\documentclass[8.5pt,twoside,twocolumn]{article}
\usepackage[super,sort&compress,comma]{natbib}
\usepackage{mhchem}
\usepackage{times,mathptm}
\usepackage{sectsty}
\usepackage{balance}
\usepackage{graphicx} %eps figures can be used instead
\usepackage{dcolumn}% Align table columns on decimal point
\usepackage{lastpage}
\usepackage{amssymb}
\usepackage{bm}
\usepackage{floatrow}
\usepackage{subfig}
\usepackage{colordvi}
\usepackage[format=plain,justification=raggedright,singlelinecheck=false,font=small,labelfont=bf,labelsep=space]{caption}
\usepackage{fancyhdr}
\usepackage{dcolumn}% Align table columns on decimal point
\usepackage[dvipsnames]{xcolor}

\newcommand{\paddyspeaks}[1]{{\color{black} #1}}
\newcommand{\azaimaspeaks}[1]{{\color{black} #1}}
\newcommand{\rlj}[1]{{\color{black}#1}}
\newcommand{\ft}[1]{{\color{black} #1}}
\newcommand{\AR}[1]{{\color{black} #1}}
\newcommand{\JH}[1]{{\color{black} #1}}

\newcommand{\phic}{\phi_{\rm c}}

\begin{document}
\thispagestyle{plain}
\fancypagestyle{plain}{
\renewcommand{\headrulewidth}{1pt}}
\renewcommand{\thefootnote}{\fnsymbol{footnote}}
\renewcommand\footnoterule{\vspace*{1pt}%
\hrule width 3.4in height 0.4pt \vspace*{5pt}}
\setcounter{secnumdepth}{5}

\makeatletter
\def\subsubsection{\@startsection{subsubsection}{3}{10pt}{-1.25ex
plus -1ex minus -.1ex}{0ex plus 0ex}{\normalsize\bf}}
\def\paragraph{\@startsection{paragraph}{4}{10pt}{-1.25ex plus
-1ex minus -.1ex}{0ex plus 0ex}{\normalsize\textit}}
\renewcommand\@biblabel[1]{#1}
\renewcommand\@makefntext[1]%
{\noindent\makebox[0pt][r]{\@thefnmark\,}#1}
\makeatother
\renewcommand{\figurename}{\small{Fig.}~}
\sectionfont{\large}
\subsectionfont{\normalsize}

\fancyfoot{}
\fancyfoot[RO]{\footnotesize{\sffamily{1--\pageref{LastPage}
~\textbar  \hspace{2pt}\thepage}}}
\fancyfoot[LE]{\footnotesize{\sffamily{\thepage~\textbar\hspace{3.45cm}
1--\pageref{LastPage}}}}
\fancyhead{}
\renewcommand{\headrulewidth}{1pt}
\renewcommand{\footrulewidth}{1pt}
\setlength{\arrayrulewidth}{1pt}
\setlength{\columnsep}{6.5mm}
\setlength\bibsep{1pt}

\twocolumn[
  \begin{@twocolumnfalse}
%\noindent\LARGE{\textbf{Gelation-enhanced sedimentation induced by confinement}}
\noindent\LARGE{\textbf{Effects of vertical confinement on gelation and sedimentation of colloids}}
\vspace{0.6cm}

\noindent\large{\textbf{%
Azaima Razali,\textit{$^{a,b,c}$}
Christopher J. Fullerton,\textit{$^{d,e}$}
Francesco Turci,\textit{$^{a,b}$}
James E. Hallett,\textit{$^{a,b}$}
Robert L. Jack,\textit{$^{d}$}
and
C. Patrick Royall\textit{$^{a,b,f,g}$}}}
\vspace{0.5cm}

\noindent\textit{\small{\textbf{Received Xth XXXXXXXXXX 20XX,
Accepted Xth XXXXXXXXX 20XX\newline
This draft: \today\newline
First published on the web Xth XXXXXXXXXX 200X}}}

\noindent \textbf{\small{DOI: 10.1039/b000000x}}
\vspace{0.6cm}
%Please do not change this text.

\noindent \normalsize{%
We consider the sedimentation of a colloidal gel under confinement in the direction of gravity. The confinement allows us to  compare directly experiments and computer simulations, for the same system size \paddyspeaks{in the vertical direction}. The \paddyspeaks{confinement} %reduced system size 
also leads to \emph{qualitatively} different behaviour compared to bulk systems: in large systems gelation suppresses sedimentation, but for small systems sedimentation is \emph{enhanced} relative to non-gelling suspensions \paddyspeaks{, although the rate of sedimentation is reduced when the strength of the attraction between the colloids is strong}. We map interaction parameters between a model experimental system (observed in real space) and computer simulations. Remarkably, we find that \paddyspeaks{when simulating the system using} Brownian dynamics %simulations 
in which hydrodynamic interactions between the particles are neglected,
% exhibit 
\paddyspeaks{we find that}
sedimentation \paddyspeaks{occurs} on the same timescale as \paddyspeaks{the} %an 
experiments, however the thickness of the ``arms'' of the gel is rather larger in the experiments, compared with the simulations. An analysis of local structure in the simulations showed similar behaviour to gelation in the absence of gravity.
}
\vspace{0.5cm}
 \end{@twocolumnfalse}
  ]

%\pacs{64.70.pv:colloids}
% 64.70.pv:  colloids

\footnotetext{\textit{$^{a}$~H.H. Wills Physics Laboratory, University of Bristol, Bristol, BS8 1TL, UK}}
\footnotetext{\textit{$^{b}$~Centre for Nanoscience and Quantum Information, Tyndall Avenue, Bristol, BS8 1FD, UK}}
\footnotetext{\textit{$^{c}$~Kulliyyah of Science, International Islamic University Malaysia, Jalan Istana, Bandar Indera Mahkota, 25200 Kuantan, Pahang, Malaysia}}
\footnotetext{\textit{$^{d}$~Department of Physics, University of Bath, Bath, BA2 7AY, UK}}
\footnotetext{\textit{$^{e}$~Laboratoire Charles Coulomb, UMR 5221, Universit\'e Montpellier, Montpellier, France}}
\footnotetext{\textit{$^{f}$~School of Chemistry, University of Bristol, Bristol, BS8 1TS, UK}}
\footnotetext{\textit{$^{g}$~Department of Chemical Engineering, Kyoto University, Kyoto 615-8510, Japan}}

\section{Introduction}

Non-equilibrium colloidal systems in gravitational fields display rich and challenging \paddyspeaks{behaviour} \cite{piazza2012}. Even the simplest colloidal system, hard spheres, exhibits a range of phenomena when the force of gravity is unleashed  \cite{russel}, due to the coupling between gravity, chemical potential \cite{piazza1993,royall2005,buzzaccaro2007,leocmach2010}, and solvent-mediated hydrodynamic interactions between the particles\cite{segre1997,segre2001nature,padding2004,wysocki2009,monchoJorda2010,wysocki2010}. \paddyspeaks{Even without gravity, }adding attractions between the colloids leads to very rich behaviour in quiescent systems \cite{poon2002,piazza2012}.  In particular, spinodal demixing can lead to a network of particles~\cite{verhaegh1997, tanaka1999colloid,manley2005spinodal,lu2008,zaccarelli2008} which undergoes dynamical arrest \paddyspeaks{--- a gel} \cite{testard2011,chaudhuri2016}. The effective attractions in these colloidal systems are induced by the addition of non-absorbing polymer. The result is a mixture of three important components --- colloids, polymers and solvent --- whose equilibrium properties can be derived from an effective one-component system of colloids with attractive interactions\JH{, where the} interaction strength is determined by the polymer concentration \cite{dijkstra2000,likos2001}.

The interplay of phase separation (which may be arrested) and sedimentation can result in novel structure-dynamical correlations~\cite{poon2002,piazza2012,buzzaccaro2007,leocmach2010,zia2014}. Among the most intriguing behaviour is that of gelation under gravity. In bulk systems, gelation typically suppresses sedimentation. This is because gelation (in the colloidal systems we consider) corresponds to the formation of a network of arrested material with finite \paddyspeaks{yield stress} %zero-shear viscosity 
\cite{zaccarelli2007,poon2002,coniglio2004, ramos2005}.  This network can then support \paddyspeaks{its own} weight, suppressing sedimentation. Gels are therefore used extensively to extend the shelf-life of many products which would otherwise sediment \cite{manley2005, macbean2009}. Under some conditions the gel can persist for years \cite{ruzicka2010}, if the self-generated or gravitational stress is weaker than the yield stress \cite{tanaka2013fara}, but gels very often undergo sedimentation \cite{piazza2014,starrs2002,buzzaccaro2012}. This is a poorly understood phenomenon and can sometimes be sudden in its onset --- so-called delayed collapse \cite{bartlett2012}. \paddyspeaks{In such delayed collapse, very little change in the macroscopic properties of the system is seen for some time, which is comparable to the timescales we consider here. Then a change occurs and the system begins to sediment on a timescale of $10^5$ particle diffusion times or more \cite{bartlett2012}. }

Here we take a radical departure from previous work in the field. Hitherto, large experimental systems have been considered, where the particles are at least $10^5$ times smaller than any linear dimension of the system, so there may be $10^{16}$ \paddyspeaks{or more} particles in the system~\cite{poon2002,bartlett2012}. The associated experimental timescales \paddyspeaks{for sedimentation} are \paddyspeaks{at least} of the order of $10^5$ diffusion times. Treating such large systems in a theoretical fashion is, at present, only possible with approximate approaches which \paddyspeaks{impose} a one-dimensional solution to the height profile such as ``batch settling'' \cite{russel} and dynamic density functional theory \cite{royall2007,schmidt2008}. To the best of our knowledge such theoretical approaches have not been extended to consider systems which undergo gelation \paddyspeaks{and in any case, the applicability to an inhmogenous materials such a gel is at least questionable.}  This leaves computer simulation as a means to treat the problem of sedimenting gels, but the timescales (up to years) and the macroscopic system sizes are not accessible \paddyspeaks{to} direct simulation.

\begin{figure}[tb]
\centering
\includegraphics[width=8cm]{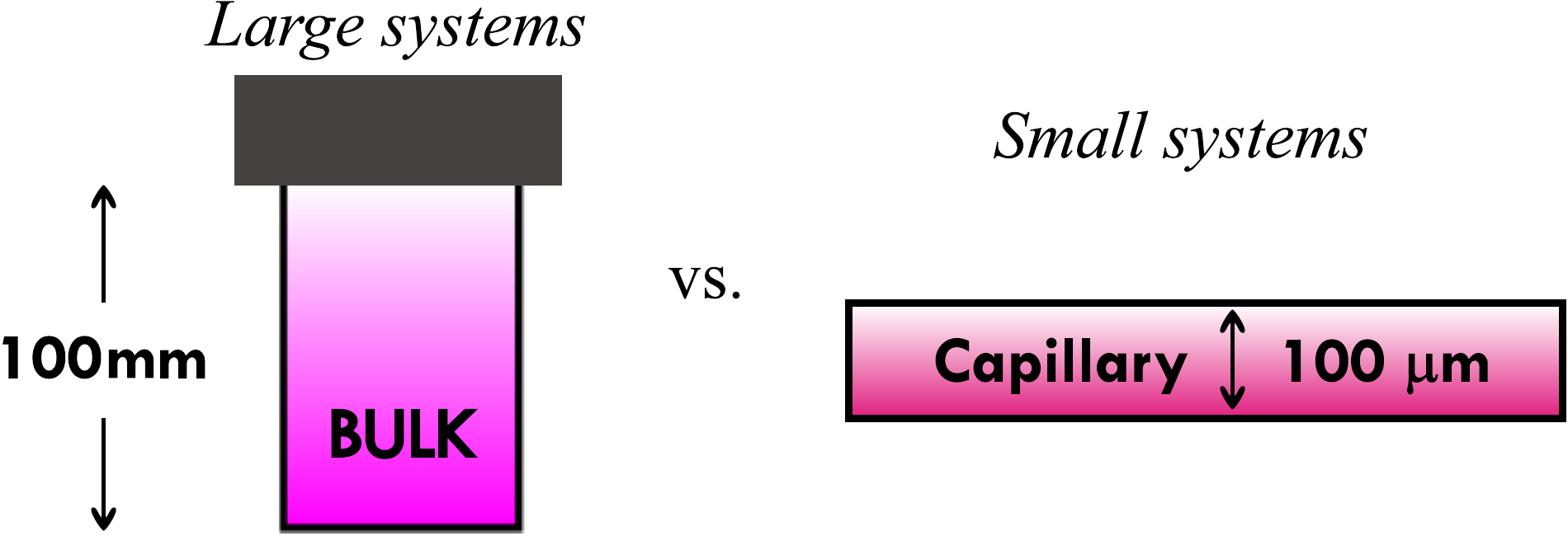} 
\caption{A sketch showing the difference between typical experimental systems in previous work~\cite{poon2002,bartlett2012}, and the system described here.}
\label{figBulkCapil} 
\end{figure}

However, it is possible to conduct experiments in much smaller systems, glass capillaries. Figure \ref{figBulkCapil} shows the difference between bulk systems (as reported previously~\cite{poon2002,bartlett2012}) and the system size used in this work. Here the relevant linear dimension (the height) is of order 100 particle diameters which is amenable to computer simulation. Such small systems thus offer a testbed by which simulation may be compared with experiment.  We employ Brownian dynamics simulations in which solvent-mediated hydrodynamic effects are included only at the one-body level (that is, Stokes drag), and hydrodynamic interactions between the particles are neglected.  Such interactions can have significant effects in sedimentation \cite{russel,wysocki2009,monchoJorda2010,wysocki2010, perez2015} and in gelation \cite{furukawa2010,royall2015}.  However, capturing them in simulations limits the accessible time scales and system sizes. \JH{Furthermore} %, and the 
Peclet numbers in these experiments are small, which we expect to reduce effects of hydrodynamic interactions.  Hence, we compare the experiments with Brownian dynamics simulations, which are simple and computationally relatively inexpensive.

Remarkably, we find semi-quantitative agreement between experiment and \paddyspeaks{Brownian dynamics} simulation. Moreover, both reveal that sedimentation in such small systems is profoundly different from that in large systems. There, gelation inhibits sedimentation, \paddyspeaks{and is used} %where it finds use 
in prolonging the shelf-life of many products. Here in small systems quite the opposite behaviour is found: \emph{gelation enhances %accelerates 
sedimentation}.

\ft{Our physical picture is the following: \paddyspeaks{in the absence of phase separation, }a bulk system of \paddyspeaks{(repulsive)} colloids under the action of gravity would attain a sedimentation profile characterised by the gravitational length $\lambda_g$; the same system, vertically confined in a capillary of length comparable to $\lambda_g$ would show \paddyspeaks{an almost constant} %no measurable 
profile. However, when  we introduce polymers into the mixture and form a gel, the collapse is \textit{slowed down} for the bulk systems while we observe that it is \textit{promoted} in the small system.  }

\ft{The sedimentation behaviour of bulk gels  has been previously extensively studied \cite{bartlett2012,harich2016} and it is known to be characterised by an initial delayed collapse followed by a slow settling that can take 
60 hours\cite{bartlett2012}. Here we focus on the observation of sedimentation in vertically confined gels, measuring \paddyspeaks{the way in which} the interaction strength influences the time evolution in experiments which is reproduced \paddyspeaks{in} simulations. From the simulations, we also obtain local structural information which helps to explain the different sedimentation behaviour for different interaction strengths.}

This article is organised as follows. In section \ref{sectionMethods}, we describe our methodology by first defining the model system and interaction potential (in section \ref{sectionModelAndInteractionPotential}) used in our experiments (in section \ref{sectionExperiment}), while details of the simulation model are given in section \ref{sectionSimulation}. \paddyspeaks{We then describe how our simulations are mapped to the experimental model system in section \ref{sectionMappingBetweenExperimentAndSimulation}.}
Next, in section \ref{sectionResults} we report how the phase behaviour for our colloid-polymer system, sedimentation dynamics and interface of collapsing gels evolve in time for different interaction strengths. Then analysis of structures formed during the sedimentation process is documented in section \ref{sectionStructural}. Finally, we conclude our discussions in section \ref{sectionConclusions}.

\section{Methods}
\label{sectionMethods}

\subsection{Model and interaction potential}
\label{sectionModelAndInteractionPotential}

\ft{For} polymers \ft{that} are much smaller than the colloids, the resulting mixture can be described by an Asakura-Oosawa (AO) model, which treats the polymer molecules as an ideal gas with hard interactions with the 
colloids\cite{vrij1976,dijkstra1999,dijkstra2000,taffs2010}. \ft{The AO \paddyspeaks{\emph{effective}}  interaction potential \paddyspeaks{between two colloids} can be written as:}

\begin{equation} 
u_\mathrm{AO} (r) 
 = \begin{cases} 
 \infty & \mbox{for } 
 r < \sigma \\ 
-\frac{k_\mathrm{B}T \pi \,(2R_{\rm g})^3 z_{\mathrm{p}}} {6}  \frac {(1+q)^3}{q^3} & \\
\times  \left[ 1 - \frac {3r}{2(1+q)\sigma} + \frac {r^3}{2(1+q)^3 \sigma^3} \right]  & \mbox{for } 
\sigma \leq r < \sigma + \sigma_{\mathrm{p}} \\ 
0 & \mbox{for } 
r \ge \sigma+\sigma_{\mathrm{p}}
\end{cases} 
\label{eqAO}
\end{equation}

\noindent
\azaimaspeaks{where the fugacity $z_{\mathrm{p}}$ is equal to the number density $\rho_{p}$ of ideal polymers in \ft{a} reservoir at the same chemical potential as the colloid-polymer mixture.} The result is an effective interaction between the colloids of range $q\sigma$ and well-depth \paddyspeaks{$u_{\rm AO}^{\rm min}$. For our parameters, Eq. \ref{eqAO} is expected to be highly accurate. For $q \leq 0.1547$ it is formally correct \cite{dijkstra1999,dijkstra2000}. However, for larger size ratios up to 0.25, the higher-order fluid structure is very well represented indeed, compared to the full Asakura-Oosawa model with explicit polymer \cite{taffs2010}.}
\paddyspeaks{We express the strength of the effective colloid-colloid interaction by the well depth:}

\begin{equation} 
u_{\rm AO}(\sigma) = u_{\rm AO}^{\rm min} = q^2 k_{\rm B} T \rho_{\rm p} \sigma^3 \frac{\pi (3+2q)}{12}.  
\end{equation}

\subsection{Experiment}
\label{sectionExperiment}

The experimental system is sterically stabilised polymethylmethacrylate (PMMA) with a diameter $\sigma =$ 460 nm suspended in cis-decalin. The colloidal polydispersity is approximately $4\%$ as determined with static light scattering. \paddyspeaks{Although hard spheres with a polydispersity of 4\% crystallise, the higher volume fraction of crystals formed in attractive systems \cite{lekkerkerker1992} means that there is more sensitivity to polydispersity. Indeed 4\% can be sufficient to greatly reduce crystallisation \cite{royall2012}.} A  colloidal gel was obtained by adding non-adsorbing polystyrene polymer with molecular weight $M_{w} = 3.46 \times 10^{6}$, leading to a polymer-colloid size ratio of $q=2R_{\rm g}/\sigma=0.3$, where $R_{\rm g}=67$ \paddyspeaks{nm} is the \paddyspeaks{estimated} polymer radius of gyration, \paddyspeaks{see section \ref{sectionMappingBetweenExperimentAndSimulation}}.

\paddyspeaks{For our parameters the gravitational length is }$\lambda_{g} = {6\mathrm{k}_{\mathrm{B}}}{T}/(g\delta\rho\sigma^{3})= 27.1 \mu{\mathrm{m}}$: here $\delta\rho$ is the density difference between the PMMA and the solvent, so $\lambda_{\rm g}$ is the height associated with a change of $k_{B}T$ in gravitational potential energy of a colloidal particle.  The P\'{e}clet number for sedimentation is then $Pe= \sigma/(2\lambda_{g}) = 8.51\times 10^{-3}$.

\paddyspeaks{As our unit of time, we use the Brownian time which we define as the typical time for a free colloidal particle to diffuse a distance comparable with its radius: $\tau_{B}=\sigma^{2}/24D=0.0317 {\rm s}$, where $D ={\mathrm{k}_{\mathrm{B}}}T/(3\pi\eta\sigma)$ is the Stokes-Einsten diffusion constant, in which $\eta$ is the solvent viscosity.}

Each sample was transferred into a 100 $\mu$m capillary and sealed with epoxy resin. \paddyspeaks{The manufacturing tolerance of these capillaries is around 10\%.} We allowed the resin to set prior to imaging and data was taken after 5 minutes. The imaging of a $z$-stack of the entire capillary height was done using time-resolved confocal microscopy (Leica SP5). For each data set, the $z$-stack images were taken at intervals of approximately 8 minutes, for a duration of 20 hours. \AR{The height of the capillaries were determined from the sample images in $xy$ and $yz$ planes. The top of the capillary is determined from the particles visibly stuck to the glass capillary walls, which are evident in our work as shown in Fig.\ref{figSedimentProfile}a. Then, this observation is continued in z-axis before the appearance of complete dark space to be the bottom of the capillary.} When reporting experimental data, we use the so-called polymer reservoir representation where the polymer concentration in the reservoir is related to that in the experiment by Widom particle insertion \cite{lekkerkerker1992}. The colloid volume fraction for each sample is extracted from the intensity measurements of the images obtained using the confocal microscope following \cite{leocmach2010}. These measurements were calibrated against homogeneous samples of known volume fraction, where there is a linear dependence of the measured intensity against colloid volume fraction.

\ft{We collect data at colloid volume fractions $\phic$ in the interval [0.1, 0.35] in order to \paddyspeaks{determine} the phase diagram of the model. When describing the sedimentation behaviour of the system as a function of the different interaction strengths induced by the different polymer concentrations, we focus on a single volume fraction $\phic=0.2$.}

\begin{figure*}[tb]
\centering
\includegraphics[width=12cm]{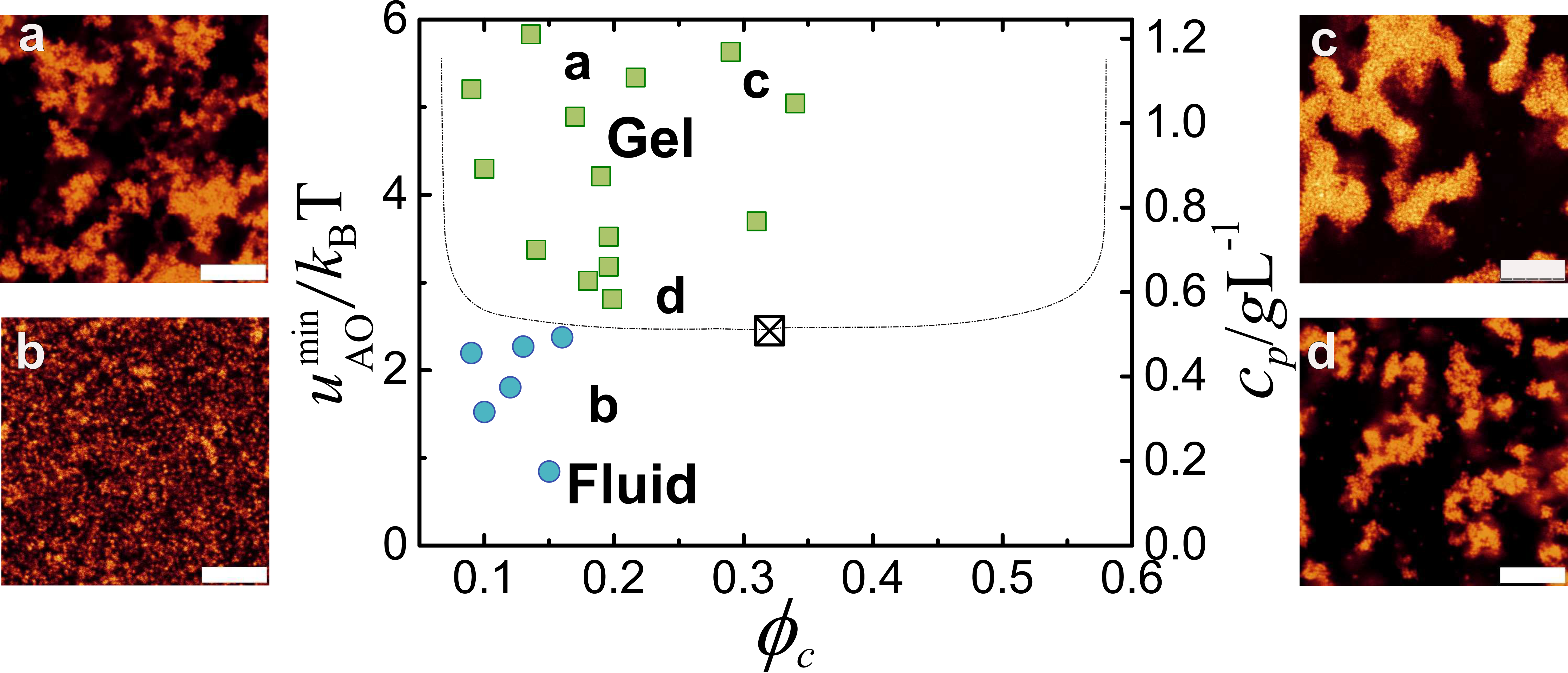} 
\caption{Summary of the states observed in the experimental colloid-polymer mixture with $q=0.3$, as a function of colloid volume fraction $\phic$ and attractive interaction strength $u_\mathrm{AO}^\mathrm{min}$, \paddyspeaks{and polymer concentration (see section \ref{sectionMappingBetweenExperimentAndSimulation})}.  Green squares indicate gels and blue circles indicate homogeneous fluids.  The $\boxtimes$ is the critical point determined based on the reduced second virial coefficient $B_{2}^{*}$ and critical isochore estimated from the literature \cite{loverso2006,royall2007nphys,taylor2012}. \azaimaspeaks{The approximate position of the spinodal is indicated by the dashed line. \ft{Two dimensional snapshots of the system illustrate} the phase behaviour observed at different points in the phase diagram:} \ft{(a) low density gel at low $\phic$ and high polymer density; (b) low polymer concentration leading to a non-percolating cluster phase; (c) high $\phic$ and polymer concentration resulting in a coarse gel network; (d) phase coexistence between fluid and gel close to the spinodal line.} 
\paddyspeaks{\JH{Sc}ale bars represent 10 $\mu$m.}
}
\label{figPhase} 
\end{figure*}

\subsection{Simulation}
\label{sectionSimulation}

\textit{Parameters and interactions. --- }
As a simple way to mimic the experimental polydispersity (whose primary effect is to suppress crystallisation), we simulate a binary mixture of particles, with equal numbers of each species.  The diameters of the two species are $\sigma_{\rm AA}=1.04\sigma$ and $\sigma_{\rm BB}=0.96\sigma$.  We consider a total of $N=60,000$ particles in a simulation box of size $L\times L \times L_z$ with $L=28.025\sigma$ and $L_z=200\sigma$, so that the volume fraction is $\phic=0.2$, as in the experiment.  All particles have mass $m$.  The boundary conditions are periodic in the $x$ and $y$ directions: there are walls at $z=0$ and $z=L_z$ that are described in detail below. The sample height $L_z$ is comparable with the dimension of the capillary used in the experiment, and the lateral dimension $L$ is comparable with the range over which experimental data was taken.
 
As a proxy for the AO potential between the colloids, we use a Morse potential
\begin{equation}
u_\mathrm{mor}(r) =  \epsilon_\mathrm{mor} \left[ \mathrm{e}^{ -2\alpha(r-\sigma_{ij})}  - 2\mathrm{e}^{-\alpha(r-\sigma_{ij})}   \right],  
\label{eqMorse}
\end{equation}

\noindent 
where $\epsilon_{\rm mor}$ is the depth of the attractive well, $\alpha$ sets the attraction range, and $\sigma_{ij}$ is the position of the minimum of the interaction between particles of species $i$ and $j$ (which depends on the particle type).  This potential accurately reproduces the behaviour of the AO system, including its higher-order local structure~\cite{taffs2010}.  In contrast to the AO potential \paddyspeaks{[Eq. \ref{eqAO}]}, the Morse potential is continuous,
\paddyspeaks{ which is convenient for simulation.} We take $\alpha = 25.0 \sigma^{-1}$ following \cite{taffs2010} and we use an additive mixing rule $\sigma_{\mathrm{AB}} = (\sigma_{\mathrm{AA}} + \sigma_{\mathrm{BB}})/2=\sigma$.  The reduced well-depth $\epsilon_{\rm mor}/(k_{\rm B}T)$ is varied between $1.0$ and $30.0$.

The particles move in an external potential that includes the effects of gravity and of the confinement by the capillary.  The gravitational potential energy of a particle at height $z$ is $E_{\rm g}(z)=zk_{\rm B}T/\lambda_{\rm g}$ with $\lambda_{\rm g}=60\sigma$, \paddyspeaks{similar to} the experiment.  The system is confined vertically by walls that are represented (for simplicity) by truncated and shifted Lennard-Jones potentials, as $u_\mathrm{wp}(\Delta z) = 4 \epsilon_\mathrm{wp}[(\sigma_\mathrm{wp}/\Delta z)^{12}-(\sigma_\mathrm{wp}/\Delta z)^{6}]$ where $\Delta z$ is the distance of the particle from the wall.  The range of the potential is $\sigma_{\mathrm{wp}} = 0.125\sigma$, comparable with the range of the Morse potential and the well-depth is $\epsilon_\mathrm{wp}=2\epsilon_{\rm mor}$.  The top wall (at $z=L_z)$ is purely repulsive, so the potential is truncated and shifted at its minimum.  The bottom wall (at $z=0$) accounts for depletion interactions between colloids and the wall, and is truncated and shifted at $r=2.4\sigma_{\rm wp}$. \paddyspeaks{The %choice of 
wall interaction behaviour was chosen to match the experiments.}

\textit{Dynamics and timescales. --- }
Langevin (or Brownian) dynamics are implemented using the LAMMPS package~\cite{lammps}.  Particles have positions $\bm{r}_i$ and velocities $\bm{v}_i$ and the velocities evolve in time as 
\begin{equation}
m \frac{\mathrm{d}}{\mathrm{d}t} \bm{v}_i = - \nabla_i V - \gamma \bm{v}_i + \sqrt{2\gamma k_{\rm B}T} \bm{\xi}_i
\end{equation}
where $V$ is the total potential energy (including contributions from particle interactions, gravity, and the confining walls), while $\gamma$ is a friction constant and $\bm{\xi}_i$ a random noise force.  The friction constant sets the time scale for the decay of velocity correlations as $\tau_{\rm d}=m/\gamma$.

There are a number of different time scales relevant for Langevin dynamics.  As well as $\tau_{\rm d}$, there is a time scale $\tau_0=\sqrt{m\sigma^2/k_{\rm B}T}$ that is independent of the damping and sets the scale for particle velocities.  Hence $\tau_0$ is the natural time unit within the LAMMPS implementation.  For colloids, the physical situation corresponds to an overdamped limit $\tau_{\rm d}\ll \tau_0$.  Here we take $\tau_{\rm d}/\tau_0=0.1$, which is small enough to give the right qualitative behaviour --- stronger damping would give a more accurate description but requires a more expensive numerical integration.  The integration time step is $0.001\tau_0$.  The single-particle diffusion constant is $D_0=k_{\rm B}T/\gamma$ so the Brownian time is $\tau_{\rm B}=\sigma^2/(24D)=\gamma\sigma^2/(24k_{\rm B}T)=\tau_{\rm 0}^2/(24\tau_{\rm d})$, which for the parameters specified above corresponds to approximately $420$ integration time steps.

\textit{Preparation of initial conditions. --- }
The system is initialised in a well-mixed state, to mimic the experimental conditions.  To achieve this, both the interparticle interactions and the interactions with the wall are truncated and shifted at their minima to achieve purely repulsive interactions.  Particles are initialised in random positions, and a conjugate gradient minimisation (without gravitational forces) is used to remove particles that are overlapping.  Then, the system is thermalised (still with purely repulsive interactions and without any gravitational forces) by evolving it for a time $50\tau_0$, leading to a homogeneous fluid configuration.  These configurations are then used as initial conditions for the main simulations (including gravity and attractive interactions) for which results are shown below. \rlj{All simulation results are averaged over 3 independent trajectories. Since the systems are fairly large, the fluctuations between trajectories are small.}

\subsection{Mapping between experiment and simulation}
\label{sectionMappingBetweenExperimentAndSimulation}

To match the state points between the Morse potential used in simulation and the (approximate) Asakura-Oosawa interactions within the experiment, we used the extended law of corresponding states \paddyspeaks{introduced} by Noro and Frenkel~\cite{noro2000}. Identical well-depths and reduced second virial coefficients $B_{2}^{*}$ (Eq.~\ref{eqB2star}) are required in order to map the state points between simulations and experiments, where
\begin{equation} 
B_{2}^{*} = \frac{B_{2}}{\frac{2}{3}\pi\sigma^{3}_\mathrm{eff}} .
\label{eqB2star}
\end{equation}
Here $B_2= 2 \pi \int_{0}^{\infty} \! [1 -\mathrm{ exp }(- \beta u)] r^2 \, \mathrm{d}r$ is the second virial coefficient and $\sigma_{\mathrm{eff}}$ is the effective diameter of a particle~\cite{hansen}. For the AO potential, $\sigma_{\mathrm{eff}}=\sigma$. For the Morse potential [Eq. \ref{eqMorse}], the effective diameter is fractionally smaller than $\sigma$, but this effect is very small for our parameters (around 1\% of the diameter) and is neglected. For the simulation results reported here, we calculated the value of $B_2^*$ associated with the relevant Morse potential, and then calculated the well-depth that would give the same value of $B_2^*$ for an AO potential with $q=0.3$.  In the following, simulation results are labelled by these effective AO well-depths, which are indicated by $u_{\rm AO}^{\rm min}$.  These effective well-depths are comparable with (but do differ from) the well-depths $\epsilon_{\rm mor}$ of the associated Morse potential.

\paddyspeaks{Accurate determination of the polymer radius of gyration is notoriously hard, with typical measurement errors around 10 \%. Alas, given that effective colloid-colloid interaction depends on the \emph{cube} of the radius of gyration, we have found that mapping to simulation and predictions (specifically that the reduced second virial coefficient $B_2^*\approx{-1.5}$ at criticality) proves a more accurate means to estimate the radius of gyration \cite{royall2007,taylor2012}. By equating the second virial coefficient \paddyspeaks{such that $B_2^*={-1.5}$ via Eq. \ref{eqAO}}, we arrive at $R_g=67$ nm, which is compatible (\emph{i.e.} within an error of 10\%) with literature data \cite{berry1966}. This corresponds to a polymer overlap concentration of $4.56 \,{\rm {gL}}^{-1}$.}

\section{Results}
\label{sectionResults}

\begin{figure}[tb]
\centering
\includegraphics[width=75mm,keepaspectratio]{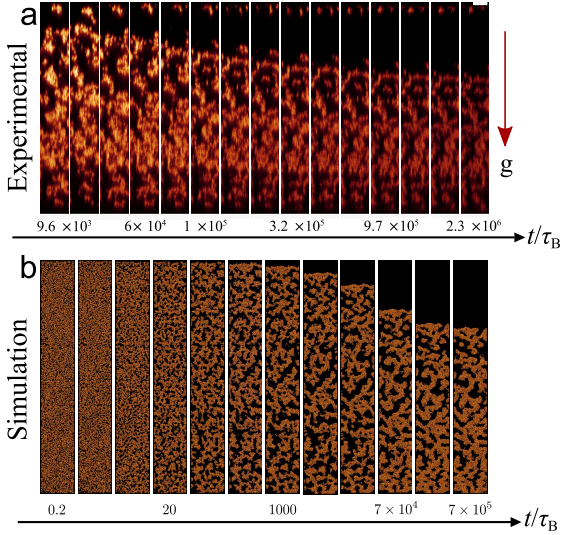} 
\caption{Time-sequence of sedimenting gels captured from (a) experiment with $u_\mathrm{AO}^\mathrm{min}=\AR{7.0}$ $k_\mathrm{B}T$ and (b) simulation corresponding to $u_\mathrm{AO}^\mathrm{min}=7.1$ $k_\mathrm{B}T$. The scale bar in (a) corresponds to 7.5 $\mu$m.  The snapshots of the experimental and simulation systems show regions of comparable size (measured in units of the colloid particle diameter $\sigma$).
}
\label{figSedimentProfile} 
\end{figure}

\subsection{Experimental phase behaviour}
\label{sectionPhaseBehaviour}

The \ft{phase diagram} of the experimental system, as a function of \paddyspeaks{well depth and polymer concentration $c_{\rm p}$} and colloid volume fraction, $\phic$ \paddyspeaks{is summarised in Figure ~\ref{figPhase}.} This \ft{diagram} is \ft{representative of}  colloid-polymer mixtures with size ratio $q=0.3$. One expects a liquid-vapor critical point in this system whose position is determined from the extended law of corresponding states \cite{noro2000}, shown here at $B_2^*=-1.5$. \paddyspeaks{In our experiments, we find that criticality occurs at a polymer volume fraction (in the so-called experimental representation) $0.56$ gL$^{-1}$.}

The critical isochore is estimated from the literature \cite{loverso2006,royall2007nphys,taylor2012}. \azaimaspeaks{The dashed line is an \ft{indicative} spinodal line.} \ft{As illustrated in the snapshots in \paddyspeaks{Fig. \ref{figPhase}}\JH{ }%the figure,  
 the system explores cluster phases (point \textbf{b}) or gel phases (points \textbf{a},\textbf{c},\textbf{d}) of different nature depending on the concentration of polymers and the colloid volume fraction: thin networks at low $\phic$ (\textbf{a}) or close to the spinodal (\textbf{d}); much coarser networks at high $\phic$ and high polymer concentration (\textbf{c}).}  

\begin{figure}[!htb]
\centering
\includegraphics[width=8.5cm,keepaspectratio]{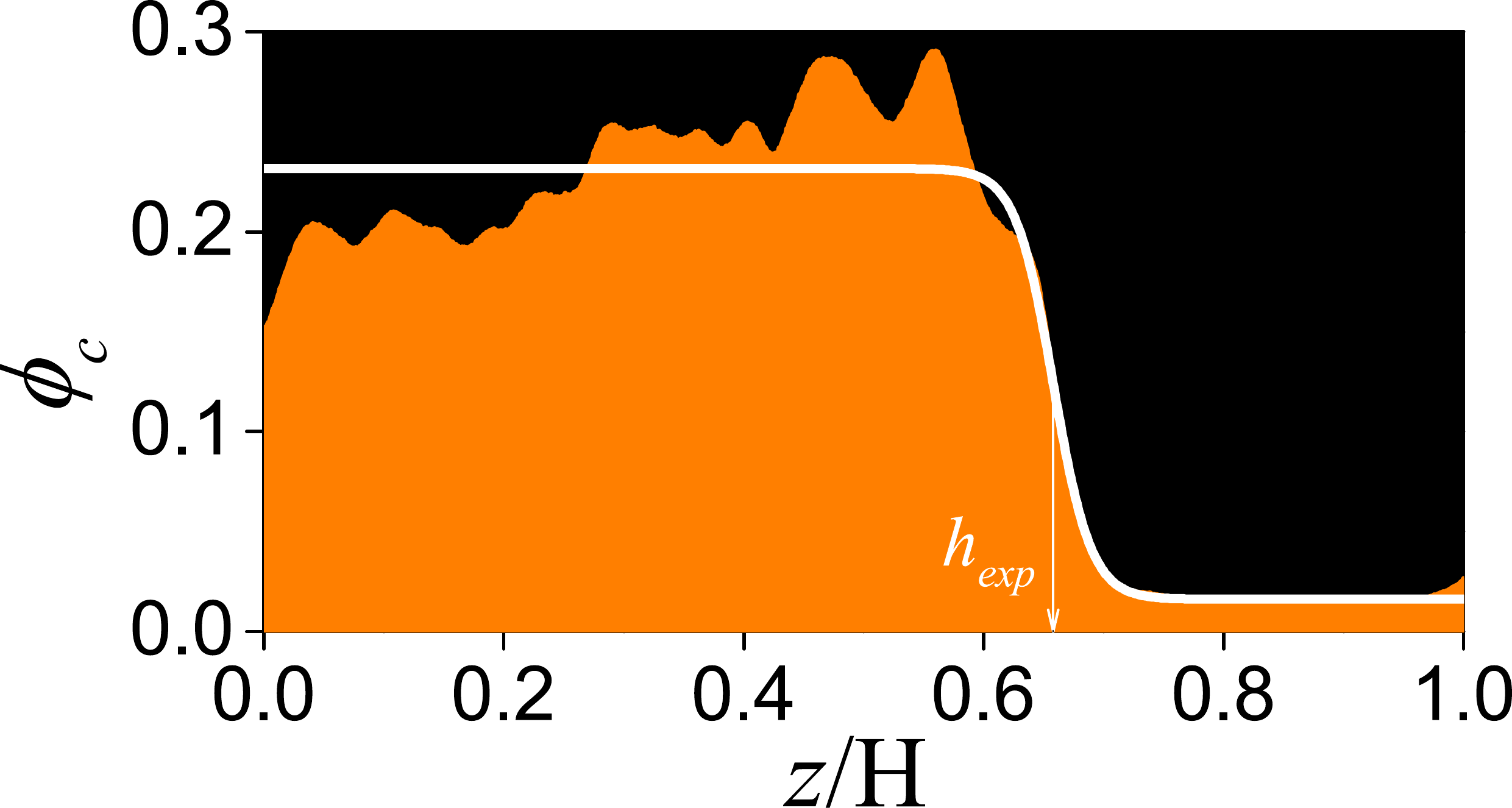} 
\caption{Sedimentation profile $\phic (z)$ for an experimental system with $u_\mathrm{AO}^\mathrm{min}=\AR{3.2}$ $k_\mathrm{B}$T at time $t=2.2\times10^{5}\, \tau_{\mathrm{B}}$. \AR{The orange area represents the average lateral packing fraction as estimated from the intensity in the xz-direction of the sample. The white line is a fit according to Eq. \ref{eqtanh}. } }
\label{figHisto} 
\end{figure}

\begin{figure*}[t]
\centering
\includegraphics[width=17.4cm]{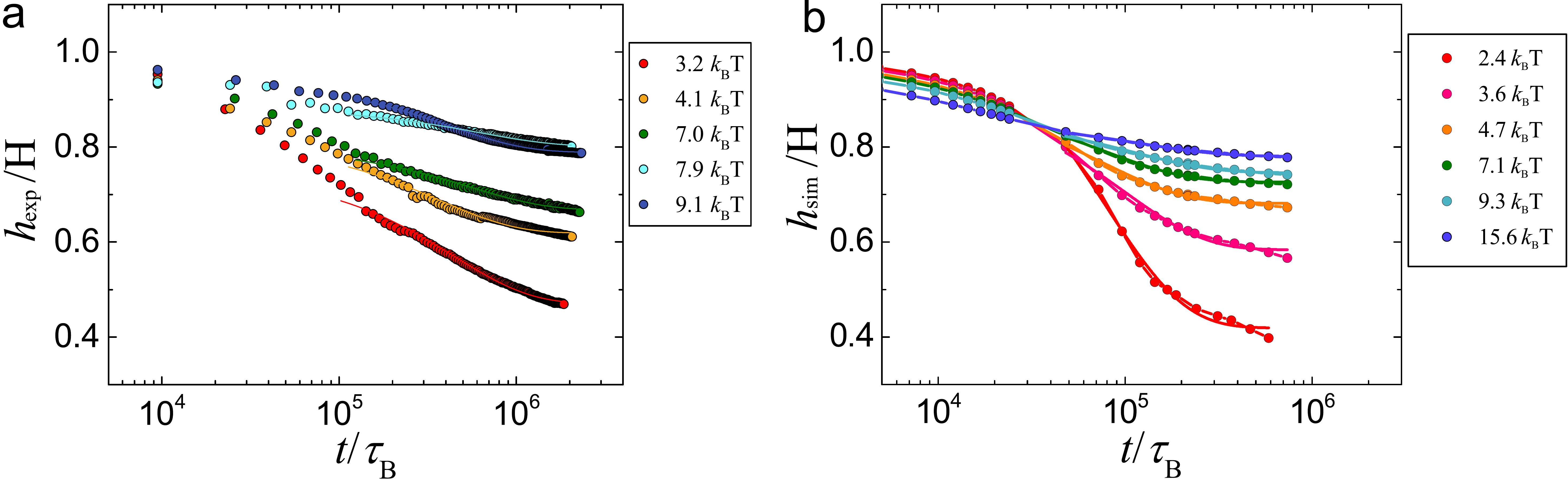} 
\caption{\AR{Figure shows \paddyspeaks{the} interface height normalised to the height \paddyspeaks{of the system}.} Gel/vapour ``interface'' height plotted as a function of time.
This height is estimated by fitting the function in (\ref{eqtanh}) to a histogram of colloid density against height. Results from experiments are shown in (a); simulation results are shown in (b). 
}
\label{figInterfaceAndBonds} 
\end{figure*}

In \ft{our} experimental samples, we see dynamically arrested gels for polymer concentrations higher than that \paddyspeaks{required} for criticality. There is no sign of colloidal liquid-gas phase separation \paddyspeaks{(either stable or metastable)}, presumably because the short range of the interaction results in dynamical arrest before phase separation can be completed. The polydispersity of the system prevents crystallisation on these time scales \cite{zaccarelli2009}.

\subsection{Global sedimentation dynamics}
\label{sectionGlobalSedimentationDynamics}

\ft{In order} to analyse the time-evolution of the \ft{colloid-polymer mixture}, we first consider the sedimentation of the system as a whole for both simulation and experiment (\ft{see} Fig.~\ref{figSedimentProfile}). At early times, one observes gelation as the formation of a percolating network of particles. At later times, particles can detach from the arms of the gel and diffuse through the solvent \paddyspeaks{or move along the ``surface'' region between the ``arms'' of the gel and the solvent \cite{royall2008,zhang2013}} : this effect induces restructuring of the gel network, and eventual collapse \cite{kilfoil2003,bartlett2012,buzzaccaro2012,zhang2013,secchi2014}. Here the dynamics of the collapsing gel is recorded by taking 3-dimensional (3d) images in the confocal microscope which span the entire capillary at different times. \paddyspeaks{Figure}~\ref{figSedimentProfile}(a) shows a sequence of such confocal images as time progresse\ft{s} for a system with $u_\mathrm{AO}^\mathrm{min}=\AR{7.0}$ $k_\mathrm{B}T$.  From the confocal images, it is evident that the gel network is initially distributed  throughout the whole capillary before falling under gravity at later times. The same qualitative behaviour is shown in the simulation data in Fig. \ref{figSedimentProfile}(b).

\ft{We determine the time evolution of the height of the collapsing gel} by plotting the local colloid volume fraction $\phic(z)$  as a function of height $z$, for each configuration in the trajectory as shown in Fig. \ref{figHisto}. From the histogram, the \ft{gel-gas} ``interface'' is obtained by fitting a hyperbolic tangent to $\phic(z)$, as
\begin{equation}
\phic(z) = \phi_0 +\delta \phi \, \tanh\left(\frac{h-z}{\xi}\right).
\label{eqtanh}
\end{equation}
Here $\phi_0$ is the mean volume fraction in the regime we are fitting and $\delta \phi$ controls the change in volume fraction across the interface. There are two fitting parameters, the height of the gel $h$, and the interfacial width $\xi$.

The fitting parameter $h$ is the height of the gel-vapour interface, which we plot in Fig. \ref{figInterfaceAndBonds}. We normalise by the total height of the system\azaimaspeaks{, $H$,} as the tolerance of the capillaries used in the experiments leads to small changes (less than 5\% in the value of $H$), thus we plot $h(t)/H$. Remarkably, the experiments and Brownian dynamics simulations exhibit sedimentation on a \paddyspeaks{comparable} timescale, and the degree of collapse is similar, although the experiments exhibit a more gradual collapse on a \paddyspeaks{somewhat} 
longer timescale than the simulations.

\begin{figure}[b]
\centering
\includegraphics[width=6cm,keepaspectratio]{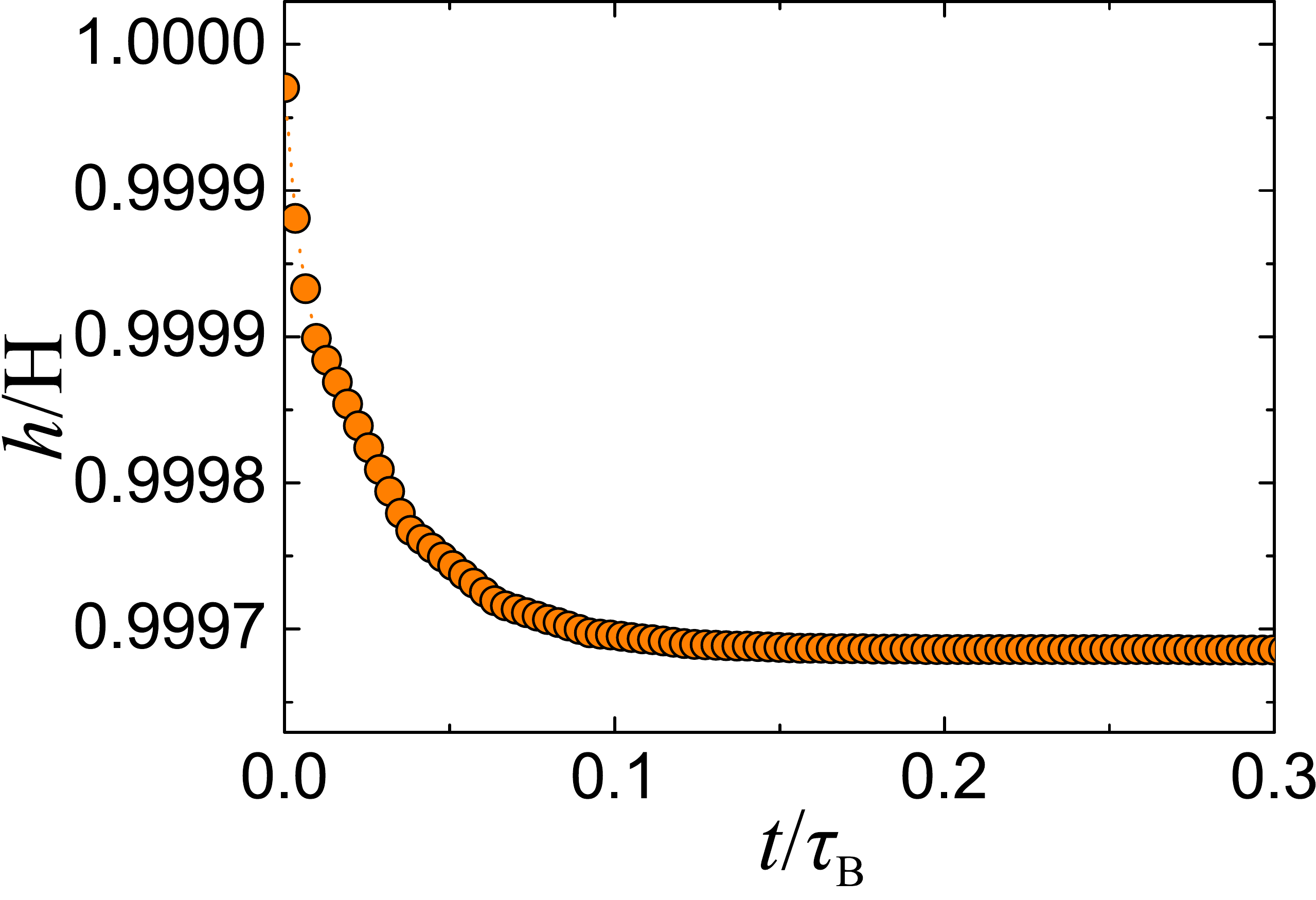} 
\caption{Sedimentation of the colloidal system in the absence of polymer. Here the P\'{e}clet number $Pe = 0.01$. Data determined from batch sedimentation \cite{russel}.}
\label{figBatch} 
\end{figure}

In the case where there are no attractive forces and the system does not form a gel, we show in Fig. \ref{figBatch} that, for our parameters, the sedimentation is negligible.  To obtain this result, we consider \emph{batch sedimentation}~\cite{russel} of hard spheres for the same capillary height and a P\'{e}clet number $Pe = 0.01$, \paddyspeaks{comparable to that of} the experimental system. The rather small change in height shows that there is little or no significant sedimentation in a colloidal system without any polymer: we also verified this fact using simulations in the regime where attractions between particles are too weak to observe gelation.

In order to estimate a timescale for the sedimentation $\tau_\mathrm{sed}$, we heuristically fit the time-evolution of the interface height with an exponential decay,
\begin{equation}
h(t)=h_{t\rightarrow \infty} +h_\mathrm{drop} \mathrm{e}^{-t/\tau_\mathrm{sed}}
\label{eqExp}
\end{equation}
where $h_{t\rightarrow \infty}$ is the interface height at long times and $h_\mathrm{drop}=h(t=0)-h_{t\rightarrow \infty}$ is the amount by which the gel-vapour interface is estimated to fall {at long times}.

\begin{figure}[tb]
\centering
\includegraphics[width=6.5cm,keepaspectratio]{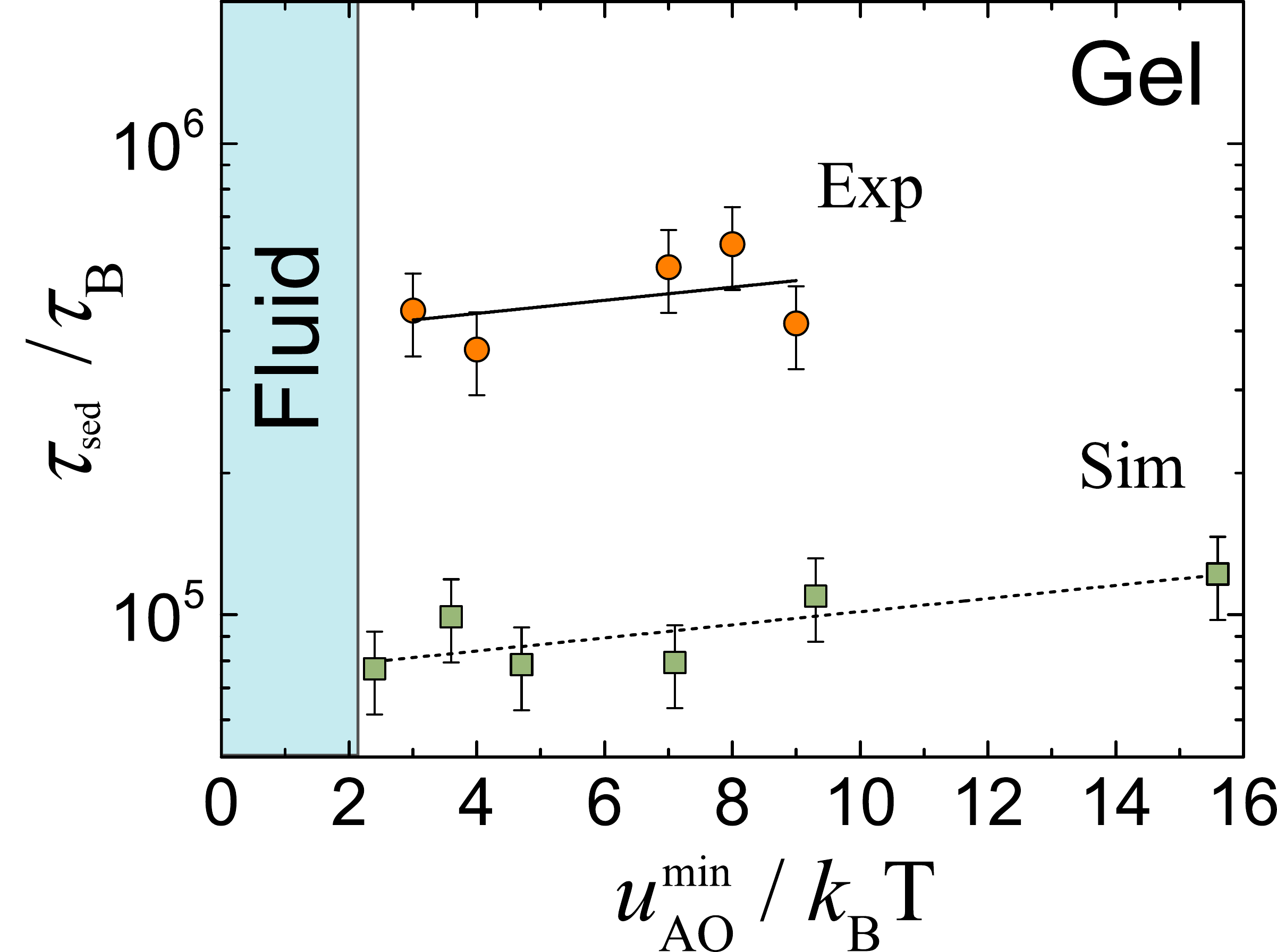} 
\caption{\AR{Sedimentation timescale $\tau_\mathrm{sed}$ as a function of interaction strength. Here we show both experimental and simulation data.} 
\paddyspeaks{Lines are fitted as described in the text.}
}
\label{figDecay} 
\end{figure}

Fits according to Eq.~\ref{eqExp} are shown in Fig. \ref{figInterfaceAndBonds}. We emphasise that this choice of fit is heuristic, and that the time-dependence of $h(t)$ is more complex than this simple exponential form, particularly for long times. Indeed we expect that $h_{t\rightarrow \infty}$ may overestimate the interface height at long times, in the case of further sedimentation on scales beyond those we access here \cite{piazza2012}. 
\paddyspeaks{We note that }\ft{Eq.~\ref{eqExp} fits the simulations better than the experiments,} suggesting some difference in the mechanism of sedimentation between experiment and simulation. Note also that the exponential fits are more accurate when the attractive interactions are stronger and the amount of sedimentation is less. \rlj{The initial sedimentation in the gels \paddyspeaks{with stronger interactions} is rapid compared to weaker gels because the condensation is faster while the coarsening is slower. The crossing point shown in the figure \ft{is believed to be purely} coincidental.}

\paddyspeaks{In Fig.~\ref{figDecay} we show} the sedimentation timescale $\tau_\mathrm{sed}$ extracted from the fits. 
\paddyspeaks{The dashed line through the simulation data is a straight line fit (in the linear-log representation of Fig. \ref{figDecay}). The solid line through the experimental data has the same slope as the fit to the simulation data, but its intercept is fitted to the experimental data.} \paddyspeaks{We find that the characteristic time of the experiments $\tau_\mathrm{sed}$ is} approximately \paddyspeaks{three times} longer in the experiments compared to the simulations. \ft{Both experiments and simulations show that as we increase the interaction strength the sedimentation timescales undergo a small increase (a factor two or three).}  Comparison with observations of bulk systems, where the interaction strength has a profound impact on the sedimentation timescale, especially in the case of delayed collapse \cite{bartlett2012}, suggests that there may be a fundamental difference in mechanism between these confined systems and bulk measurements.  Certainly the behaviour shown in Fig.~\ref{figBatch} would be very different in bulk systems, where the system height is much greater than the gravitational length, so batch settling under gravity would lead to significant sedimentation even in the absence of attractive forces between colloids.

The discrepancy in time scales between simulation and experiments in Fig.~\ref{figDecay} has several possible origins. \ft{We exclude from these our choice of interaction potential, because} the Morse potential used in the simulation has previously been shown to capture quite accurately the behaviour of this class of experimental system~\cite{royall2008,royall2008aip,royall2015}, and the AO model also matches such experiments~\cite{royall2007}. Moreoever, the Morse and AO systems are also very similar to each other~\cite{taffs2010}, so we expect this aspect of the simulations to be reliable. \ft{Also, we exclude our choice to mimic in the simulation} the effects of continuous polydispersity in the experimental system by the use of a binary mixture: this %coarse 
approximation (in the absence of significant crystallisation) seems unlikely to affect sedimentation time scales in \ft{the} way \ft{that we observe}. One \ft{possible origin of the discrepancy} is that hydrodynamic interactions are important: these have been shown to have \paddyspeaks{considerable} influence in the time-evolution of gels in the absence of sedimentation \cite{furukawa2010,royall2015}. \ft{A second} possibility %, as we discuss below, 
is that the well-mixed initial conditions used in the simulations do not match the state of the colloidal suspensions at the beginning of the experiments. \paddyspeaks{Finally we note that the simulations consider a finite periodic system while the experiments consider a small part of a much larger system. Of course it would be desirable to consider larger systems but this is not feasible due to the associated computational cost.}

\subsection{Structural behaviour upon coarsening}
\label{sectionStructural}

Having analysed the height of the gel as a function of time, we now analyse the structure within the gel itself.
This analysis takes two forms. First we consider the thickness of the network, which coarsens over time, as also happens in systems where sedimentation does not play an important role \cite{zhang2013, zia2014,testard2014}. To do this we determine the \emph{chord length} \cite{testard2011}. We then perform a local structural analysis at the particle level on the simulation data using the topological cluster classification \cite{malins2013tcc} and common neighbour analysis \cite{honeycutt1987}.

{\begin{figure*}[!htb]
\begin{center}
\includegraphics[scale=0.17]{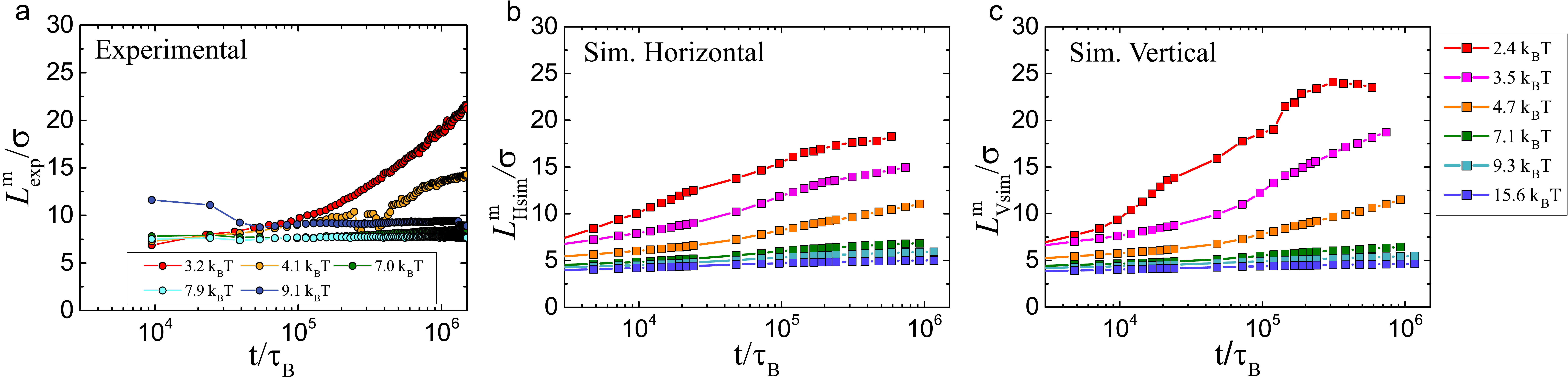}
\caption{Average chord length (mass weighted) measured 
(a) in the horizontal direction in the experimental system \AR{at state points correspond\JH{ing} to Fig.~\ref{figInterfaceAndBonds}}.
(b) in the horizontal direction (perpendicular to the direction of gravity) (simulation), 
(c) in the vertical direction (the direction in which gravity acts)  (simulation).
\paddyspeaks{Legend on right pertains to (b) and (c).}
}
\label{figChords}
\end{center}
\end{figure*}

\textit{Chord length. --- }
    \rlj{It is useful to estimate the typical size of the arms of the gel (Fig.~\ref{figPhase}).  We achieve this by measuring a chord length, following \cite{testard2011}.  In order to identify the arms of the gel, it is useful to measure the local density in the system.  For a given point $\bm{R}_\alpha$, we define a (non-normalised) measure of local density as $n_\alpha = \sum_i f(|\bm{r}_i-\bm{R}_\alpha|)$.  
\paddyspeaks{Here $f(r)={\rm e}^{-r^2/\ell^2}$ is a (non-normalised) Gaussian smoothing function, with $\ell=0.25\sigma$.}  %If the cell is within one of the arms of the gel then we expect $n_\alpha$ to be  large; if the cell is in the vapour then $n_\alpha$ should be small.  
%We identify each cell as either dense (gel) or dilute (vapour) by comparing $n_\alpha$ with a cutoff $n^*=0.3$.
The quantity $n_\alpha$ is large if point $\alpha$ is inside an arm of the gel, and small if the point is in the \paddyspeaks{colloid-poor phase}.  We take a threshold $n=0.3$, so $R_\alpha$ is in the gel if $n_\alpha>0.3$ and in the sol if $n_\alpha<0.3$ (The distribution of $n$ is bimodal so results depend weakly on this threshold). We carry out this analysis for a 3d cubic grid of points with spacing $0.5\sigma$. A chord is a straight line that cuts through an arm of the gel.  Chords may have any direction. As a representative sample, we identify chords that run along the $x$, $y$, and $z$ directions.  We achieve this by running through the cubic grid (along the lattice axes) and identifying all sets of contiguous cells for which $n>0.3$. We record the length of each chord. }

Chords measured in the $x$ and $y$ directions are equivalent (as gravity acts in the $z$ direction only), but we separate horizontal chords (aligned along the $x$ and $y$ directions) and vertical chords (aligned long $z$).  To estimate the typical size of a horizontal chord, imagine choosing a particle at random and measuring the chord containing that particle.  If the length of the $j$th horizontal chord is $H_j$ then the average length of a horizontal chord chosen in this way is
\begin{equation}
L^{\rm m}_{\rm H} = \frac{\sum_j H_j^2 }{\sum_j H_j}
\label{eqChord}
\end{equation}
where the superscript `m' indicates that the average is \emph{mass-weighted}. (That is, this average could equivalently be estimated by choosing particles at random and measuring the associated chords.  On the other hand, averaging the length of a randomly chosen chord would give a different result.  The mass-weighted average focusses attention on the chords which contain the majority of the particles and avoids numerical artefacts associated with large numbers of small chords.)

This typical chord length is shown in Fig. \ref{figChords} for both experimental and simulation data. We see that \paddyspeaks{at long times} the chord lengths for the experimental systems are significantly larger than those in the simulations.  However, except for this difference in overall scale, the time-evolution in both experiment and simulation appears similar.

There are several possibilities for this observation. The first is that the time-evolution is somehow different between the experiments and simulations, perhaps due to hydrodynamic interactions \cite{furukawa2010,royall2015}.  Alternatively, the lateral size of the simulation box could influence the size of the networks formed.  Previous simulation studies have emphasised the need for large systems in order to avoid finite size effects on the gel structure \cite{testard2011,testard2014}. (Note however that the lateral system size $L\approx 28\sigma$ is comparable with the range over which experimental data was taken.)

\paddyspeaks{Concerning the response of the system to increasing attraction strength, we see that in the gel regime, increasing attraction strength appears to lead to a suppression of domain growth in both experiments and simulations. This is in keeping with the literature \cite{teece2011,zhang2013,testard2014}. }

\subsection{Local structural analysis}

Gelation is accompanied by significant changes in local structure~\cite{royall2008}. We therefore probe the local structure in our \paddyspeaks{simulations of} sedimenting gels, for which we consider two methods of analysis.  The first is the Topological Cluster Classification (TCC) \cite{malins2013tcc} and the second is a common neighbour analysis (CNA)\cite{honeycutt1987}.  These measurements were performed as a function of the height within the gel but we found little vertical variation in the relative population of local structures (despite the density difference in the sedimentation profiles).  In the following, we therefore plot the {population of local structures averaged across the whole system.}

\begin{figure*}[!htb]
\begin{center}
\includegraphics[scale=0.138]{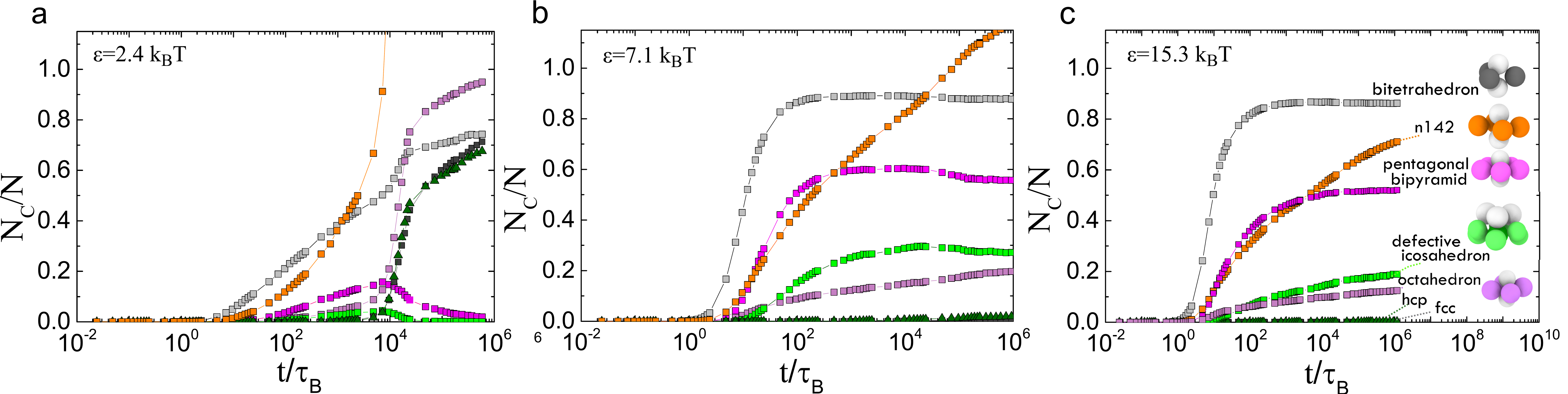}
\caption{Time-evolution of the local structure in simulations.  We consider three state points for which the effective AO well depths are $\epsilon_\mathrm{eff}=2.4, 7.1$ and $15.3$ $k_\mathrm{B}T$. 
The `142' clusters are detected by the common neighbour analysis, other structures are found with the topological cluster classification (TCC). For TCC clusters, we show the fraction of particles ($N_C/N$) that participate in at least one cluster of the relevant type.  Hence one clearly has $N_C/N\leq 1$.  For 142 clusters, $N_C/N$ is the average number of 142-bonds in which a particle participates, so one may have $N_C/N>1$ (indeed for a perfect fcc crystal one would have $N_C/N=12$).
}
\label{figCluster}
\end{center}
\end{figure*}

\textit{Topological cluster classification. --- }
In this structural analysis, isolated clusters of particles were identified that represent energy minima of the Morse potential  (with $\alpha= 25.0/\sigma$).  Then, bond networks of the simulated gel structures were calculated using a modified Voronoi construction, and all 3, 4 and 5 membered shortest-path rings \azaimaspeaks{were identified within these bond networks. We then set a tolerance for asymmetry in the 4 membered rings, denoted $f_c$. We set this to 0.85, consistent with previous work \cite{malins2013tcc}}. Then, local structures within the bond network that are topologically equivalent to the original energy minima were identified and enumerated. The clusters identified using the TCC are illustrated in Fig.~\ref{figCluster}, as are the proportions of particles that participate in clusters of each type. Since particles may be identified in more than one type of structure, the total across different types may exceed one.

\textit{Common neighbour analysis. --- }
The common neighbour analysis (CNA)\cite{honeycutt1987} offers a way to classify bonds. A bonded pair is classified based on how many mutual neighbours they share, and how these mutual neighbours are bonded.
Of primary interest are 142 bonds, which are found in large numbers in both the HCP and FCC crystals \paddyspeaks{and thus are here interpreted as a crystal precursor}.  (These 142-bonds~\cite{klotsa2011} include both the 1421 and 1422 bonds of the original scheme~\cite{honeycutt1987}). Figure \ref{figCluster} shows the average number of 142 bonds that a particle participates in for different well depths. Here, a particle participates in a 142 bond if it is one of the two particles forming the central bonded pair.

In Fig.~\ref{figCluster} we plot \paddyspeaks{the populations of} a number of local structures known to be important in gelation \cite{royall2008,royall2008aip,royall2012}.
The data reveals a number of observations. The first is that all three state points exhibit similarities in their behaviour. At short times, there are few structures. Upon condensation, (the first stage of gelation), local structures form, beginning with the 5-membered bitetrahedron. This is similar to previous work in quiescent (non-sedimenting) systems \cite{royall2012}, and we note that the tetrahedron is the simplex for spheres in 3d, so its prevalance at early times is expected. Again, similar to previous studies \cite{royall2012,royall2015}, we see a tendency to the 10-membered defective icosahedron at longer times.

Upon weak quenching (see Fig.~\ref{figCluster}(a)), we find a considerable degree of crystallisation at longer times, as observed previously in related systems \cite{klotsa2011}. Moreoever, the appearance of crystalline order as measured by the TCC occurs up to an order of magnitude later in time then the emergence of crystalline 142-bonds as measured by the CNA. While these 142-bonds are associated with crystallisation, they represent a lower degree of local order than the 13 particle clusters that are identified as fcc/hcp  in the TCC analysis. Increasing the strength of attractions leads to a suppression of crystallisation in the timescales accessible here, consistent with previous work \cite{royall2008,royall2008aip,klotsa2011,royall2012,taylor2012}.

In summary, given the change in state parameters, the \paddyspeaks{time} evolution of \paddyspeaks{local structure of} our sedimenting gels is not markedly different to that of quiescent gels \cite{royall2012}. We note that while we expect the binary system used in these simulations to mimic the large-scale properties of the experiments, the presence of only two component types does have the potential to influence local structure (and crystallisation) when compared to the continuous polydispersity of experimental systems.  Note, however, that in contrast to both the experimental and simulation systems considered here, monodisperse systems  crystallise much more easily~\cite{royall2012,doye2005}.

\section{Conclusions}
\label{sectionConclusions}

We have carried out a combined experimental and simulation study of colloidal gels undergoing sedimentation. The vertical confinement of these systems profoundly affects their sedimentation behaviour. \ft{In particular we observed that for confined colloidal systems for which the gravitational length would not be compatible with a sedimentation profile, the addition of polymers and the resulting gelation \emph{induced} sedimentation \paddyspeaks{in systems which essentially do not sediment in the absence of gelation}. \paddyspeaks{Quite unlike} bulk systems, no delayed collapse is observed. }

Our Brownian dynamics simulations provide a reasonable description of the time-evolution of the system. This is possible due to a careful mapping of the interaction parameters between experiment and simulation.
The agreement between simulation and experiment is notable, given that the simulations do not feature hydrodynamic interactions. The major differences in behaviour between simulation and experiment are that the simulations appear to sediment \paddyspeaks{rather} faster than the experiments. Structural analysis on the dimensions of the gel network suggests that the experiments are rather coarser \paddyspeaks{at long times}.  
This may be related to some intrinsic difference in the dynamics, or to a finite size effect in the simulations, or to incomplete homogenisation of the experimental system prior to gelation.

We have also considered the local structure of the simulated gels.  We find that this is rather similar to the structural evolution found in quiescent (non-sedimenting) systems.  Recalling from Fig.~\ref{figSedimentProfile}(b) that the system clearly condenses into a percolating network before any significant sedimentation has occurred, it seems that the main changes in the local structure of the system occur on short time scales that are decoupled from sedimentation.  Figure~\ref{figCluster} is also consistent with this interpretation.

Finally, we note that simulation studies such as these might provide a basis by which coarse-grained theoretical models might be developed, which \paddyspeaks{could potentially} tackle truly macroscopic systems.  This would be valuable since macroscopic phenomena such as delayed gel collapse~\cite{bartlett2012} are not accessible in these small (confined) systems, and are therefore beyond the reach of direct simulation.  For this reason, development of such coarse-grained models would form a major step forward in the understanding and modelling of these important materials. \paddyspeaks{A most interesting outcome of such an approach would be the successful prediction of sedimentation \emph{rates} several orders of magnitude faster than those observed here, as observed in delayed collapse \cite{bartlett2012}. It is possible that such studies would be helped by larger scale simulations than those we have been able to perform here. Those we have carried out lie at the limit of our resources. We carried out smaller scale simulations with a reduced height and saw identical behaviour, save that the height was scaled according to the system size.}

\section*{Acknowledgments}

Peter Crowther is gratefully acknowledged for kind assistance with the image data analysis.
CPR acknowledges the Royal Society, the Japan Society of the Promotion of Science (JSPS) and Kyoto University SPIRITS fund for financial support and EPSRC grant code EP/H022333/1 for the provision of a confocal microscope. AR is grateful to the Malaysia's Ministry of Education (MOE) for the financial support and thanks R. Pinchaipat for the assistance in data analysis. \paddyspeaks{CPR\JH{, JEH} and FT gratefully acknowldge the European Research Council (ERC consolidator grant NANOPRS, project number 617266)}
CJF and RLJ acknowledge support from the UK Engineering and Physical Science Research Council (EPSRC) through grants EP/I003797/1 and EP/L001438/1.

\footnotesize{
\bibliography{falling.bib,extra.bib}
\bibliographystyle{rsc}
}

\end{document}